# Comment:
# A socio-technical framework for digital contact tracing

Ricardo Vinuesa[1]*, Andreas Theodorou[2], Manuela Battaglini[3] and Virginia Dignum[2]**
[1]*SimEx/FLOW, Engineering Mechanics, KTH Royal Institute of Technology, SE-100 44 Stockholm, Sweden*
[2]*Responsible AI Group, Department of Computing Science, Umeå University, SE-90358 Umeå, Sweden*
[3]*Transparent Internet, Tårup Bygade 30, DK-5370 Mesinge, Denmark*
Corresponding authors: **rvinuesa@mech.kth.se;* ***virginia@cs.umu.se*

**Abstract**
In their efforts to tackle the COVID-19 crisis, decision makers are considering the development and use of smartphone applications for contact tracing. Even though these applications differ in technology and methods, there is an increasing concern about their implications for privacy and human rights. Here we propose a framework to evaluate their suitability in terms of impact on the users, employed technology and governance methods. We illustrate its usage with three applications, and with the European Data Protection Board (EDPB) guidelines, highlighting their limitations.

**Introduction and motivation**

Around the world, policymakers and public-health experts are calling for the use of *contact-tracing applications* as means to fight COVID-19. The aim is to track those who come in contact with infected people with the aid of smartphone applications (referred to as *apps* in this paper). The basic premise behind these apps is that whenever an individual is diagnosed with the coronavirus, every person who had possibly been near that infected individual during the period in which they were contagious is notified and told to either self-quarantine or request COVID-19 testing. Proponents of this approach point out that the apps can help to stop the chain of transmission in order to control the outbreak without the need of a full lockdown.

Many tracking apps are being introduced via a fast-tracked development circle, often paid with public resources, with very limited socio-economic impact assessment and concern for fundamental rights and values such as *fairness* and *inclusion* [1,2]. It is therefore important to critically consider the actual usefulness, necessity and effectiveness of the apps, as well as their impact on the broader social system, including our fundamental rights and freedoms, considering that these apps set a precedent for future use of similar invasive technologies, even after the COVID-19 crisis.

Central to the decision to employ such apps, is the question of their effectiveness for the containment of spread. Some studies suggest that a penetration of 60% of the population [3] would be needed, while others show that as much as 80-90% penetration would be required[1]. Current implementations in different countries do not seem to achieve penetration higher than 25%. Furthermore, as indicated by the World Health Organization (WHO)[2], discrimination may arise against users, or against those that for various reasons are not able to use the app, and history has shown that the interplay of surveillance and epidemiology can unfortunately also lead to threats and violence against certain groups.

In order to aid in the evaluation of these apps, here we propose a framework which is adapted from the one used in our recent article on artificial intelligence (AI) and the Sustainable Development Goals (SDGs) [2], to evaluate current approaches and concerns related to development, deployment and usage of tracking apps.

**Evaluation criteria and example cases**

The proposed evaluation framework is based on a total of 19 criteria, divided into the three following categories: *Impact on the citizens*, *Technology* and *Governance*. These criteria are derived from different regulations and guidance documents [4,5] and from the concerns raised by experts[3]. Each criteria is measured on a scale from 0 to 2 as discussed next.

---

[1] See the work by the Agent-based Social Simulation of the Coronavirus Crisis research group (ASSOCC): https://simassocc.org/scenario-effect-of-tracktrace-apps/
[2] See the WHO report: https://www.who.int/violence_injury_prevention/violence/world_report/en/summary_en.pdf

Impact on the citizens

1. **Respecting fundamental rights of individuals:** This includes the rights to safety, health, non-discrimination and freedom of association (2). Unclear information/only partially respecting these rights (1), or not respecting them (0) are not adequate.
2. **Privacy and data protection:** Data collection should be compliant with the General Data Protection Regulation (GDPR) [6] and respect the privacy of the individual. A Data Protection Impact Assessment (DPIA) must be carried out before the deployment of any contact-tracing system. The purpose of the app and the mechanisms to assess its usage need to be clearly defined. All these requirements should be fulfilled (2), whereas fulfilling them only partially (1) or not at all (0) are not adequate.
3. **Transparency rights:** They include the right of users to be notified, to control their own data, transparency regarding which personal data are collected, and of explanation of app-produced output. The app should be auditable. Fulfillment of all requirements (2) is suitable, whereas fulfilling them only partially (1) or not at all (0) are not adequate.
4. **Avoid discrimination:** The app needs to prevent stigmatization due to suspected infection (2). Unclear information/measures to avoid this (1), or the lack of a plan to address this issue (0) are not adequate.
5. **Accessibility:** Possibility to be used by all regardless of demographics, language, disability, digital literacy and financial accessibility. All these requirements should be fulfilled (2), whereas addressing them only partially (1) or not at all (0) are not adequate.
6. **Education and tutorials:** Ensure that users are informed and capable of using the app correctly, including e.g. in-app help (2), or external materials, e.g. website (1). Absence (0) is not adequate.

Technology

7. **Decentralized protocol:** E.g. use of the Decentralized Privacy-Preserving Proximity Tracing (DP-3T) architecture [7]. Furthermore, the app needs to allow interoperability. Bluetooth is preferred over GPS. A fully decentralized protocol is best (2), whereas mixed (1) or completely centralized approaches are not adequate (0).
8. **Data management:** Ensure data-minimization principle, i.e. usage of local and temporary storage, and encryption, based on principles of data protection by design. Ensure that only data strictly necessary are processed. All these requirements are needed (2), whereas unclear documentation (1) or lack of compliance with all of them (0) are not adequate.
9. **Security:** User authentication to prevent risks such as access, modification, or disclosure of the data. Use unique and pseudo-random identifiers, renewed regularly and cryptographically strong. Compliance with these requirements is needed (2), whereas unclear (1) or lack of compliance (0) are not adequate.
10. **App easy to deactivate/remove:** Either through clear instructions or automatically by sunset clause (2). Unclear (1) or difficulties for removing the app and the data (0) are not adequate.
11. **Open-source code:** Participatory and multidisciplinary development, access to the code and methods used for adaptation to new knowledge on the virus (2). Open-source code without the possibility of contributing (1) is not recommended, and non open-source code is undesirable (0).

Governance

12. **Public Ownership:** Ownership by State is preferable (2), whereas Health Agency (1), a research institute (1) or a private/commercial party (0) are less adequate.
13. **Data governance should be made public:** Open data governance is preferable (2), while intermediate (1) or private/opaque settings (0) are not suitable.
14. **Use**: Downloading the app needs to be voluntary (2). Furthermore, the use of the app cannot be mandatory to access certain places (1) or otherwise be legally enforced (0).

---

[3] See for example a letter to the prime minister of the Netherlands by a group of scientists: http://allai.nl/wp-content/uploads/2020/04/Online-version-Letter-to-President-Rutte-Ministers-De-Jonge-Van-Rijn-Grapperhaus-re.-COVID-19-apps.pdf; and also a similar letter to the Spanish government: https://www.transparentinternet.com/es/transparencia/carta-al-gobierno-de-espana-covid-19/

15. **Sunset clause:** This needs to be clearly specified with a clear date and procedure (2), while unclear information (1) or the lack of such a clause (0) are not adequate.
16. **Legislation and Policy:** Clear, broader legal framework voted through parliament (2), partial governmental policy (1) whereas no policy or unknown is not desirable (0).
17. **Incidental Findings and dual-use policy:** Purposes beyond contact tracing (e.g. placing people into crime scenes, identification of behaviour patterns) are strictly prohibited (2). If not, at least a policy stating what are the other potential uses of the data collected (1) needs to be in place.
18. **Design Impact Assessment and Open Development Process:** Explicit design process, including clear description about aims and motivation, stakeholders, public consultation process and impact assessment (2). Unclear information (1) or the lack of such an assessment (0) are not adequate.
19. **Right to contest/liability.** Users need to be able to contest decisions or demand human intervention (2). Partial/unclear compliance (1) or the lack of this feature (0) are not adequate.

As an example of application of this framework, Figure 1 shows the result for three apps: Stopp Corona [8] (currently under development in Austria), NHS COVID-19 [9] (being developed in the United Kingdom) and TraceTogether [10] (which has been deployed and utilized in Singapore since March 20, 2020). In addition, we also analyze the European Data Protection Board (EDPB) guidelines [4], and assess to what extent they comply with our framework. We observe that all the apps have low scores in *Governance*, and none of them complies with criteria 15, 17 and 19, which are in our view important areas for any digital contact tracing. The EDBP guidelines provide a clause to halt the use of apps once the situation returns to 'normal'. This can be seen as vague, since 'normal' is open to interpretation considering the socio-economic changes lockdowns brought. A more clear date, unless further action is taken, would be preferred. The EDPB guidelines also require criterion 19, but they do not include any requirement regarding geotagging (relevant for criterion 17). It is also important to remark the importance of using a decentralized protocol (criterion 7), a feature which is not exhibited by the NHS COVID-19 app and it is not required by the EDPB guidelines, while TraceTogether only partly complies with it through a mixed centralized/decentralized protocol. We believe that this approach should be implemented in any digital contact-tracing app, in order to fully ensure the safety of citizen data.

## Discussion and concluding remarks

The COVID-19 pandemic is revealing two conflicting perspectives: governments need sufficient epidemiological information to manage the pandemic, whereas citizens while wanting safety are concerned about privacy, discrimination, and personal-data protection. In order to ensure that the goals from both perspectives are achieved, transparency regarding the problems associated with collection and processing of personal data is essential [11,12]. Information about which measures are in place to safeguard human rights and freedoms, and about how the app is developed and governed is a primary condition for the acceptance of these apps.

This work contributes to this goal by providing decision makers that are considering the introduction of digital contact-tracing apps a framework to assess possible alternatives in terms of societal impact, technology, and principles of responsible development and governance. The tool can also be used by individuals considering the use of these apps to assess how they align with their needs and concerns.

## Acknowledgements

R.V. acknowledges funding provided by the Swedish Research Council (VR). A.T. is funded by the European Union's Horizon 2020 research and innovation programme under grant agreement No 825619 (AI4EU project). V.D. is supported by the Wallenberg AI, Autonomous Systems and Software Program (WASP), funded by the Knut and Alice Wallenberg Foundation. The authors also acknowledge Steen Rasmussen and Loïs Vanhée for constructive discussions about the technical issues and social impact associated with digital contact-tracing apps.

## Author contributions

The study was ideated and designed by R.V., A.T., M.B. and V.D. The four authors analyzed the data and wrote the article.

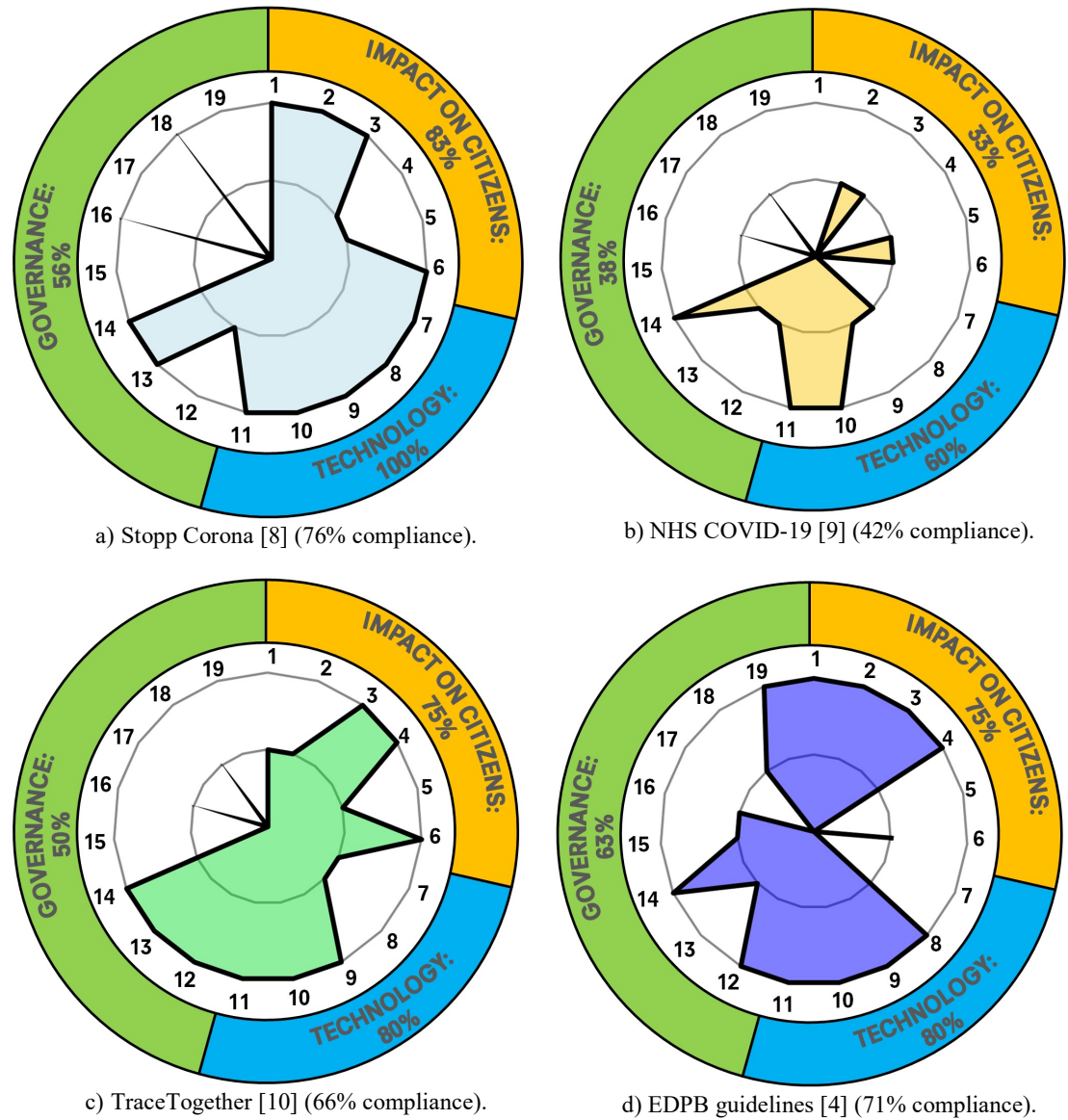

Figure 1: Application of the proposed framework to three apps and the EDPB guidelines, as indicated in each panel. The numbers represent each of the criteria, and the compliance with the criteria from the three main groups are shown in the outer circle.